# Optical read-out and control of antiferromagnetic Néel vector in altermagnets and beyond

A. V. Kimel[1], Th. Rasing[1] and B. A. Ivanov[1,2]

[1] *Radboud University, Institute for Molecules and Materials, Nijmegen, The Netherlands*

[2] *Institute of Magnetism, National Academy of Sciences and Ministry of Education and Science, 03142 Kiev, Ukraine*

**Finding methods for the most efficient and fastest detection and control of magnetic domains in antiferromagnets is presently among the main challenges of magnetic research at large. We analyse the problem of optical read-out and control of the antiferromagnetic Néel vector using symmetry analysis and the principles of equilibrium thermodynamics. Following the pioneering approach of Dzyaloshinksii, we divide all antiferromagnets in three classes. It is shown that, using the magneto-optical Faraday effect or other effects which scale linearly with the antiferromagnetic Néel vector, it is possible to distinguish antiferromagnetic domains with mutually opposite Néel vectors in two of the three classes. Symmetry properties of one of these two classes are similar to those of altermagnets. The analysis also reveals multiple mechanisms to directly excite spins with light for practically every type of antiferromagnet.**

## 1. Introduction

Traditionally, all magnetic materials are divided into two classes: ferro- and antiferromagnets. While ferromagnets are known from ancient times and characterized by a spontaneous magnetization, antiferromagnets were discovered only in the 20$^{th}$ century and often referred to as magnets with no magnetization. Such a simplified view on antiferromagnets, however, is incomplete and certainly not fully correct [1]. For example, this definition does not allow to understand how two seemingly similar antiferromagnetic transition metal oxides with identical crystal structures can have radically different magnetic properties.

For a long time antiferromagnets were a subject of purely fundamental interests, but in recent decades it became clear that antiferromagnetic materials also offer unique functionalities. For instance, in the quest for higher magnetic storage density, it was realized that antiferromagnets can be used to stabilize ferromagnetic domains [2]. In addition, similar to ferromagnets, antiferromagnets can also host magnetic domains themselves and thus can serve as a magnetic storage medium for information technologies. As the frequency of spin resonance in antiferromagnets may reach 1 THz [1], according to theory, writing magnetic bits in antiferromagnetic materials can be up to 1000 times faster than writing in their ferromagnetic counterparts. The interest in antiferromagnets has recently been fuelled by their potential to revolutionize spintronics [3,4] and magnonics [5], and thus to facilitate ultrafast processing of magnetically stored information. Hence, finding the most efficient and the fastest possible way to detect and control magnetic domains in antiferromagnets are presently among the main challenges of magnetic research at large.

As a laser pulse is practically one of the shortest stimuli in condensed matter physics, it is natural to ask if antiferromagnetic domains can be detected and controlled with the help of light. For instance, it is well known that, using the magneto-optical Faraday effect, magnetic domains in ferromagnets can be visualized with light. Can we do the same in antiferromagnets? This question has existed from the very beginning of antiferromagnetism, but in recent years it has increasingly become a subject of intense research and debates [6].

This review is inspired by the recent emergence of the field of altermagnetic materials, where magnetic order with vanishing net magnetization, typical of antiferromagnets, is accompanied by time-reversal symmetry-breaking responses and spin-polarization phenomena, typical of ferromagnets [7]. Altermagnets are especially appealing as potential sources of spin-polarized currents for antiferromagnetic spintronics. Therefore, the lion share of today's altermagnetic research is dedicated to conducting materials. Here we would like to emphasize that time-reversal symmetry-breaking responses, such as the magneto-optical Faraday effect, have been subject of fundamental research in antiferromagnetic dielectrics for more than five decades. Although the conclusions drawn from those studies can greatly spur our understanding of altermagnetism, these earlier works are often overlooked.

The present paper has three goals: (a) to formulate the rules allowing to predict if light can probe spins in antiferromagnets via the magneto-optical Faraday effect; (b) to define the main mechanisms allowing optical control of magnetism in these antiferromagnets; (c) to demonstrate that these questions can be answered without any analysis of the electronic structure of antiferromagnets, but using the symmetry approach successfully applied to antiferromagnetic dielectrics in the past [8-10].

We divide all antiferromagnets into three classes and review the earlier studies of light-spin interaction in antiferromagnets, including those which are now classified as altermagnets. We demonstrate that the appearance of time-reversal symmetry-breaking responses like the magneto-optical Faraday effect, is not a sufficient criterium to classify altermagnets. We also review peculiarities of antiferromagnetic spin dynamics and describe the main channels for optical control of spins for each of the three classes of antiferromagnets.

## 2. Antiferromagnetism

Antiferromagnets were discovered only in the 20$^{\text{th}}$ century and the discovery was initiated by the development of quantum theory of ferromagnetism. After the discovery of electron spin, it was realized that the interaction of two free electrons is spin dependent. This spin-dependent part of the electron-electron interaction is called exchange interaction. According to the model of isotropic Heisenberg exchange coupling, the Hamiltonian of the interaction is

$$\widehat{H}^{(\text{Heis})} = -J\widehat{\mathbf{S}}_1\widehat{\mathbf{S}}_2 \qquad (1),$$

where $\widehat{\mathbf{S}}_1$ and $\widehat{\mathbf{S}}_2$ are spin operators of two interacting electrons and $J$ is the exchange integral [11]. In ferromagnets, the exchange interaction favours parallel alignment of spins ($J > 0$) and thus can result in a net magnetization of the sample volume $V$: $\mathbf{M} = -g\mu_B \sum_i \mathbf{S}_i /V$, where $g$ is the Lande factor (for magnetic ions in the s-state, such as $Fe^{3+}$, $g \approx 2$) and $\mu_B$ is the absolute value of the Bohr magneton. The magnetization is an axial vector which reverses its sign under operation of time-reversal $\mathbf{M} \rightarrow -\mathbf{M}$. Soon after the discovery of the exchange interaction, it was realized that if the exchange integral is of opposite sign ($J < 0$), it favours antiparallel alignment of spins and thus results in other types of magnets called antiferromagnets and ferrimagnets, respectively.

In the simplest case, if everywhere in the material with $J < 0$, two interacting spins are equivalent, $\mathbf{M} = 0$ and such material is called an antiferromagnet. If $J < 0$, but magnetic ions with antiparallel orientation of spins are not equivalent $\mathbf{M} \neq 0$ and the material is ferrimagnetic. Such an intuitive picture is, however, oversimplified and, strictly speaking, valid only for isotropic media. The isotropic Heisenberg exchange interaction given by Eq.1 implies that the magnetic system is invariant with respect to any simultaneous rotation of all the spins in the magnet. In crystalline magnets, this is not the case and the complete Hamiltonians of magnetic systems are far more complex.

We consider a relatively simple case of an antiferromagnetic crystal, which we model as two antiferromagnetically coupled sublattices with the magnetizations $\mathbf{M}_1$ and $\mathbf{M}_2$, respectively. According to the Néel-Landau theory, it is convenient to describe the magnetic order of such an antiferromagnet using the antiferromagnetic Néel vector $\mathbf{L} = \mathbf{M}_1 - \mathbf{M}_2$. Two magnetic domains of the antiferromagnet would thus correspond to areas with mutually opposite $\mathbf{L}$. It is clear that in order to satisfy the condition $\mathbf{M} = \mathbf{M}_1 + \mathbf{M}_2 = 0$ in a broad range of external parameters (temperature, pressure, electric field, magnetic field etc), two interacting spins must be in equivalent environments. As every crystal is characterized by a set of symmetry operations that transform the crystal into itself, it is clear that we should include the consequences of symmetry in our analysis.

**An antiferromagnet must have at least one symmetry operation allowed by its crystal structure (i.e. crystal symmetry of this material in the paramagnetic phase), which transforms one magnetic sublattice into another. Such symmetry elements in antiferromagnetic crystals are called odd symmetry elements [10].**

Such a symmetry constrain is unique to antiferromagnets. Ferro- and ferrimagnets may even not have any crystal structure as in the case of amorphous alloys, for instance. In contrast to ferro- and ferrimagnetic materials, antiferromagnets are extremely sensitive to their crystal structure. Consequently, aiming to understand the physics of antiferromagnets, a symmetry analysis plays an absolutely important, indispensable role.

### 3. Phenomenology of light-spin interaction in magnets

The polarization rotation of linearly polarized light upon propagation through a magnetized medium, discovered by M. Faraday in 1845, was the very first demonstration that light can probe magnetism. The effect is now called magneto-optical Faraday effect. The magnetization $\mathbf{M}$ can be either spontaneous or induced by an external magnetic field. Therefore, qualitatively the magneto-optical Faraday effects in diamagnetic, paramagnetic and ferromagnetic materials are similar. For antiferromagnets the situation is complicated by the presence of more than one magnetic sublattice, which can result in magneto-optical effects that have no analogies in ferromagnets.

We again consider the simplest case of an antiferromagnet with just two magnetic sublattices having the magnetization $\mathbf{M}_1$ and $\mathbf{M}_2$, respectively. We will further work with two mutually orthogonal vectors $\mathbf{L} = \mathbf{M}_1 - \mathbf{M}_2$ and $\mathbf{M} = \mathbf{M}_1 + \mathbf{M}_2$ ($\mathbf{LM} \equiv 0$).

We will use the electric-dipole approximation, assuming that light-matter interactions are dominated by the interaction of the electric field of light with electric dipoles in the material. Using the first law of thermodynamics and assuming that there is no dissipation in the process of light matter interaction, i.e. the total entropy does not change, we can write an expression for the specific internal energy (internal energy per unit volume) as

$$u = u_0 + \frac{1}{2}\varepsilon_0 \varepsilon_{kl} E_k(\omega) E_l^*(\omega) \qquad (2),$$

where $u_0$ represents the part of the specific internal energy that does not depend on light, $\varepsilon_0$ is the permittivity of free space, $\varepsilon_{kl}$ is the tensor of the dielectric permittivity, $E_k$ is the $k$-th component of the electric field of light at the frequency $\omega$ and $E_l^*$ is the complex conjugate of the $l$-th component of the electric field of light. Light can probe magnetization induced by an external magnetic field **H**, a spontaneous magnetization **M** as well as an antiferromagnetic Néel vector **L**, if the dielectric permittivity is a function of these vectors.

If there is no dissipation, it can be shown that $\varepsilon_{kl} = \varepsilon_{lk}^*$ [12]. In this case, it is convenient to introduce $\varepsilon_{kl}$ as a sum of an antisymmetric $\varepsilon_{kl}^{(a)} = -\varepsilon_{lk}^{(a)}$ and a symmetric $\varepsilon_{kl}^{(s)} = \varepsilon_{lk}^{(s)}$ part

$$\varepsilon_{kl} = \varepsilon_{kl}^{(a)} + \varepsilon_{kl}^{(s)} \tag{3}.$$

It follows that $\varepsilon_{kl}^{(a)}$ results in circular birefringence and thus the magneto-optical Faraday effect, while $\varepsilon_{kl}^{(s)}$ can result in linear birefringence of the corresponding medium [13]. If there is dissipation, $\varepsilon_{kl}^{(a)}$ and $\varepsilon_{kl}^{(s)}$ additionally result in circular and linear dichroism, respectively[13]. From the expression for the specific internal energy we can obtain a thermodynamic definition of the dielectric permittivity

$$\varepsilon_{kl} = \frac{2}{\varepsilon_0} \frac{\partial^2 u}{\partial E_k^* \partial E_l} \tag{4}.$$

Applying the time reversal operation $+t \to -t$, one gets $\mathbf{H}(+t) = -\mathbf{H}(-t), \mathbf{M}(+t) = -\mathbf{M}(-t), \mathbf{L}(+t) = -\mathbf{L}(-t), E_k(\omega; +t) = E_k^*(\omega; -t)$. As during the interaction with light the total entropy of the system does not change, $u(\mathbf{H}, \mathbf{M}, \mathbf{L}; +t) = u(-\mathbf{H}, -\mathbf{M}, -\mathbf{L}; -t)$ and

$$\varepsilon_{kl}(\mathbf{H}, \mathbf{M}, \mathbf{L}; +t) = \frac{2}{\varepsilon_0} \frac{\partial^2 u(\mathbf{H},\mathbf{M},\mathbf{L};+t)}{\partial E_k^*(+t) \partial E_l(+t)} = \frac{2}{\varepsilon_0} \frac{\partial^2 u(-\mathbf{H},-\mathbf{M},-\mathbf{L};-t)}{\partial E_k(-t) \partial E_l^*(-t)} \tag{5}.$$

If the order of differentiation does not affect the outcome, i.e. if $u$ is an analytic function of $E_k$ and $E_l$, we obtain

$$\varepsilon_{kl}(\mathbf{H}, \mathbf{M}, \mathbf{L}; +t) = \varepsilon_{lk}(-\mathbf{H}, -\mathbf{M}, -\mathbf{L}; -t)$$

As the total entropy of the system does not change, the system is invariant with respect to time reversal, so that we can omit the time-variable and thus find that

$$\varepsilon_{kl}(\mathbf{H}, \mathbf{M}, \mathbf{L}) = \varepsilon_{lk}(-\mathbf{H}, -\mathbf{M}, -\mathbf{L})$$

and

$$\varepsilon_{kl}^{(a)}(\mathbf{H}, \mathbf{M}, \mathbf{L}) = \varepsilon_{lk}^{(a)}(-\mathbf{H}, -\mathbf{M}, -\mathbf{L}) = -\varepsilon_{kl}^{(a)}(-\mathbf{H}, -\mathbf{M}, -\mathbf{L}) \tag{6a},$$

$$\varepsilon_{kl}^{(s)}(\mathbf{H}, \mathbf{M}, \mathbf{L}) = \varepsilon_{lk}^{(s)}(-\mathbf{H}, -\mathbf{M}, -\mathbf{L}) = \varepsilon_{kl}^{(s)}(-\mathbf{H}, -\mathbf{M}, -\mathbf{L}) \tag{6b}.$$

We can expand $\varepsilon_{kl}^{(a)}$ and $\varepsilon_{kl}^{(s)}$ in power series of **H**, **M** and **L**. $\varepsilon_{kl}^{(a)}$ and $\varepsilon_{kl}^{(s)}$ are invariant under the operation of space inversion. **H** and **M** are axial vectors i.e. they are also invariant with respect to space inversion. At the same time, if space inversion is an odd symmetry element of an antiferromagnet, it should change the sign of **L**. Hence, one arrives at a paradox - the antiferromagnetic Néel vector **L**, defined as a difference of two axial vectors, does not satisfy the definition of either polar or axial vector. There are different ways to resolve this paradox and predict if an antiferromagnet can show magneto-optical phenomena linear in **L** [7,14,15].

In any case, aiming to write a series for $\varepsilon_{kl}^{(a)}$ and $\varepsilon_{kl}^{(s)}$ that is valid for any antiferromagnet independent of its symmetry, one realizes that terms linear with respect to the antiferromagnetic Néel vector are forbidden by symmetry, which leads to

$$\varepsilon_{kl}^{(a)} = \chi_{kln}^{(M)} M_n + \chi_{kln}^{(H)} H_n \tag{7a},$$

$$\varepsilon_{kl}^{(s)} = \varepsilon_{kl}^{(0)} + \chi_{klnp}^{(M)} M_n M_p + \chi_{klnp}^{(H)} H_n H_p + \chi_{klnp}^{(HM)} H_n M_p + \chi_{klnp}^{(L)} L_n L_p \tag{7b},$$

where $\chi_{kln}^{(M)}, \chi_{kln}^{(H)}, \chi_{klnp}^{(M)}, \chi_{klnp}^{(H)}, \chi_{klnp}^{(L)}$ are phenomenological tensors and $\varepsilon_{kl}^{(0)}$ is the part of the dielectric permittivity that does not depend on **H**, **M** or **L**. These expressions implicitly state that, assuming that light-matter interaction is of electric dipole origin and linear, optical observation of antiferromagnetic domains with mutually opposite **L** seems to be impossible. Despite this fact, the rich family of antiferromagnets provides plenty of examples showing the opposite. To understand and demonstrate this without accounting for nonlinear light-matter interactions of higher order [16], we can classify antiferromagnets in three classes with respect to their odd symmetry elements. In particular, the odd symmetry elements defining three classes of antiferromagnets are (A) translation, (B) proper or improper rotation, (C) space inversion.

According to Ref. [17], in altermagnets "the opposite-spin sublattices are connected by a real-space rotation transformation (proper or improper and symmorphic or non-symmorphic), but not connected by a translation or inversion." Hence, class (B) in our classification has similar symmetry properties as the class of altermagnets in the classification of Refs. [7,17].

### 4. Light-spin interaction in three classes of antiferromagnets

*4A. Antiferromagnets with translation as odd symmetry element*

Consider an antiferromagnet with just two magnetic sublattices (**M**$_1$ and **M**$_2$), where along a certain crystallographic direction the period of the magnetic structure is twice as large as the period of the crystal lattice **a**. Examples of such antiferromagnets are NiO and CoO, while it was shown that Cr can also be treated as an antiferromagnet with translation as the odd symmetry element [18]. In such antiferromagnets, the crystal lattice is invariant under a translation of atoms along this crystallographic direction by **a**. At the same time, this translation changes the antiferromagnetic Néel vector $\mathbf{L} \rightarrow -\mathbf{L}$, while, obviously, it does not affect $\varepsilon_{kl}^{(a)}$, $\varepsilon_{kl}^{(s)}$, **H** and **M**.

**In antiferromagnets, where translation is the odd symmetry element, optical read-out of antiferromagnetic domains with mutually opposite Néel vectors, at least in thermodynamic equilibrium, is impossible and even distinguishing domains with opposite Néel vectors does not make much sense.**

It is well known that, if one considers only the spin system of an antiferromagnet, antiferromagnetic domains are thermodynamically unfavourable. Nevertheless, they are formed in real antiferromagnets. Lattice defects play a decisive role in this process. A dislocation is one of such defects. Let us consider a screw dislocation, which can be visualized as the result of "cutting" a crystal along a plane and slipping one half across the other by a lattice vector, the halves fitting back together without leaving a defect. If the cut only goes part way through the crystal, and then slips, the boundary of the cut is a screw dislocation. Such a dislocation converts the lattice plane into a helical surface like a spiral staircase without steps. The axis of the helical surface is called dislocation line. If in a perfect crystal, regarded as a continuous medium, the distance between two infinitely close points is $d\mathbf{l}$ and as a result of a defect the distance is changed to $d\mathbf{l}'$, one can define the displacement vector as $d\mathbf{u} = d\mathbf{l}' - d\mathbf{l}$. Integration of the displacement vector around any closed curve $\Lambda$, which encloses the dislocation line, gives:

$$\oint_\Lambda d\mathbf{u} = -\mathbf{b}.$$

where $\mathbf{b}$ is called the Burgers vector of the dislocation [19]. For a screw dislocation $\mathbf{b}$ is equal to the lattice vector $\mathbf{a}$ in direction and magnitude. If translation is the odd symmetry element of an antiferromagnet, a screw dislocation thus leads to a paradoxical situation, because a turn by 360 degrees along the curve $\Lambda$ reverses the antiferromagnetic Néel vector $\mathbf{L} \to -\mathbf{L}$ [20]. It means that two areas corresponding to two domains with mutually opposite antiferromagnetic Néel vectors are not necessarily separated by a domain wall. Hence, in theory [21] a domain wall separating areas with mutually opposite antiferromagnetic Néel vectors can suddenly stop at a dislocation. Domain walls, suddenly broken presumably at crystal defects, were indeed observed experimentally in antiferromagnetic Cr [22].

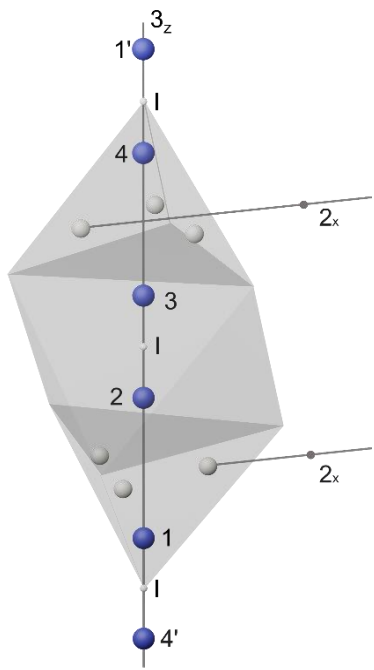

**Figure 1**. *Crystal structure of altermagnetic hematite. Blue spheres are magnetic ions, where numbers 1 (1'), 2(2'), 3(3'), 4(4') indicate the positions of the atoms corresponding to the magnetic sublattices with magnetization $\mathbf{M}_1$, $\mathbf{M}_2$, $\mathbf{M}_3$, $\mathbf{M}_4$, respectively. I indicates the inversion centre. Grey spheres are non-magnetic ions such as $O^{2-}$. The odd symmetry element here is a $2_x$ rotation axis.*

*4B. Antiferromagnets with rotation axis as odd symmetry element*

If the odd symmetry element of an antiferromagnet is a rotation axis, the situation is similar to the case of altermagnets (see Ref.17). Here we give just a couple of examples of this class of magnets, while their detailed analysis is given in Refs.8,10. We consider the crystal structure of hematite shown in Fig. 1. It is invariant with respect to the space inversion ($I$), rotations over 180-degrees around the $x$-axis ($2_x$) and rotations over 120-degrees around the z-axis ($3_z$). The spins of $Fe^{3+}$ ions in positions "1" and "4" are mutually parallel. The spins of $Fe^{3+}$ in positions "2" and "3" are also mutually parallel, but pointing in the opposite direction with respect to those in positions "1" and "4". The magnetizations of the $Fe^{3+}$ ions in the corresponding positions of the crystal are $M_1$, $M_2$, $M_3$, $M_4$, respectively. For the net magnetization we have $M = M_1 + M_2 + M_3 + M_4$. The antiferromagnetic Néel vector we define as $L = M_1 - M_2 - M_3 + M_4$.

For this antiferromagnet, the odd symmetry element is the $2_x$ rotation. In order to emphasize that this is the odd element, we will notate it as $2_x^{(-)}$. If we apply the operation of space inversion (I), it will permute $M_1$ and $M_4$ as well as $M_2$ and $M_3$. Hence, neither $M$ nor $L$ will change. Rotations around the $z$-axis will also affect $M$ and $L$ equivalently. Applying the $2_x^{(-)}$ operation of 180-degrees rotation around the $x$-axis, the $x$-projection of any axial or polar vector will not change ($2_x^{(-)} M_x = M_x$), while the $y$ and $z$ projections of the vectors must change sign ($2_x^{(-)} M_y = -M_y$, $2_x^{(-)} M_z = -M_z$). The fact that 180-degrees rotation around the $x$-axis permutes $M_1$ and $M_2$ as well as $M_3$ and $M_4$ has a dramatic effect on the magnetic properties of the medium. It means that upon 180-degrees rotation around the $x$-axis, the $x$-component of the antiferromagnetic Néel vector $L$, in contrast to the magnetization $M$, does change sign, while the $y$- and the $z$-components are not affected ($2_x^{(-)} L_x = -L_x$, $2_x^{(-)} L_y = L_y$, $2_x^{(-)} L_z = L_z$).

As it was shown by I. E. Dzyaloshinksii in his seminal work on "weak ferromagnetism" of antiferromagnets [23,24], $L_x M_y - L_y M_x$ is invariant under all symmetry operations of a hematite crystal. The same invariant can be written as $\mathbf{e}_z \cdot (\mathbf{M} \times \mathbf{L})$, where $\mathbf{e}_z$ is the unit vector in the direction of the $z$-axis. . Due to this invariant, if $M$ and $L$ are in the $xy$ plane, a purely collinear alignment of spins experiences a distortion – spin canting, which results in a weak, non-zero magnetic moment. The phenomenon is known today as the Dzyaloshinksii-Moriya interaction. Interestingly, for antiferromagnets with weak ferromagnetism, where $\mathbf{e}_z \cdot (\mathbf{M} \times \mathbf{L})$ is an invariant, rotation axis representing the odd symmetry element of the antiferromagnet is never parallel to $\mathbf{e}_z$. A detailed analysis of symmetries of antiferromagnetic crystals allowing such an invariant, and thus having spin structures affected by the Dzyaloshinksii-Moriya interaction, is given in Ref.10. Here, we would like to emphasize that in antiferromagnets with such a crystal symmetry already in the ground state, we can write $M_y = D_{yx} L_x$ and $M_x = D_{xy} L_y$, where $D_{yx}$ and $D_{xy}$ are components of the tensor of the Dzyaloshinskii-Moriyia interaction. Note that for the case of hematite $D_{xy} = -D_{yx}$ and $M_z$ is always null. It means that by symmetry $L_y$ would induce magneto-optical effects as if it was $M_x$ and $L_x$ would induce magneto-optical effects as if it was $M_y$.

For instance, if light propagates in the direction of the $y$-axis, it will experience circular birefringence and the magneto-optical Faraday effect, if $\varepsilon_{xz}^{(a)} = -\varepsilon_{zx}^{(a)} \neq 0$. In an isotropic medium, we can expect two contributions to the dielectric permittivity, which can be induced by an external magnetic field $H_y \neq 0$ and magnetization $M_y \neq 0$, respectively. In hematite, in the expression for $\varepsilon_{kl}^{(a)}$ we obtain an additional term, because the symmetry of the crystal allows to do the following substitution $M_y \to M_y + D_{yx} L_x$. Hence, the expression for the dielectric permittivity can be rewritten as

$$\varepsilon_{xz}^{(a)} = \chi_{xzy}^{(M)} M_y + \chi_{xzy}^{(H)} H_y + \chi_{xzx}^{(L)} L_x \qquad (8).$$

Since under any symmetry operation allowed by the crystal $M_y$ and $L_x$ transform similarly, the third term, linear with respect to $L_x$ and the phenomenological tensor $\chi_{xzx}^{(L)}$, appears in $\varepsilon_{xz}^{(a)}$ and results in an antiferromagnetic Faraday effect.

Note that similar derivations can also be repeated for the electric conductivity $\sigma_{kl}$, where its antisymmetric part $\sigma_{kl}^{(a)} = -\sigma_{lk}^{(a)}$ would result in a Hall effect. Hence, for hematite one can expect that

$$\sigma_{xz}^{(a)} = \varrho_{xzy}^{(H)} H_y + \varrho_{xzy}^{(M)} M_y + \varrho_{xzx}^{(L)} L_x \qquad (9).$$

where $\varrho_{xzy}^{(H)}, \varrho_{xzy}^{(M)}, \varrho_{xzx}^{(L)}$ are components of a phenomenological tensor describing the corresponding susceptibilities. The last term of the expression describes the antiferromagnetic Hall effect, which was already reported more than 40 years ago. K. B. Vlasov et al studied the Hall effect in crystals of hematite above the Morin point [25,26], where the antiferromagnet has easy-plane magnetic anisotropy with both vectors **M** and **L** in the *xy* plane. Applying an external magnetic field in the *xy*-plane and sending the current also in the *xy*-plane perpendicular to the magnetic field, the authors of Ref [25,26] measured the Hall voltage along the *z*-axis. In Fig. 2, the Hall signal first shows a rapid increase and reaches a saturation at the fields corresponding to those that turn the sample into a monodomain state. Further increase of the field does not result in any substantial changes of the Hall signal. Note that the antiferromagnetic vector in fields much smaller than the characteristic field of the exchange interaction (100 T) can be assumed to be field-independent. At the same time, the field- and magnetization-dependent terms in Eq.(9) are expected to show much more pronounced dependencies on the strength of an external magnetic field. Hence, the experimentally observed Hall voltage appears to be dominated by the field-independent antiferromagnetic term $\varrho_{xzx}^{(L)} L_x$.

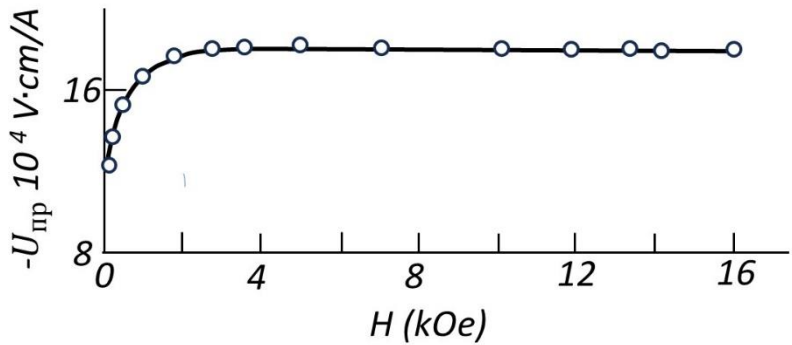

**Figure** 2. *Antiferromagnetic/altermagnetic Hall effect in a hematite crystal. Normalized Hall voltage as function of magnetic field H. The image was digitized from [25]. Units were translated from Russian.*

Hematite $\alpha$-$Fe_2O_3$ is not the only uniaxial antiferromagnet, where the odd symmetry element is two-fold rotation around an axis, which is not the main axis of the crystal Other antiferromagnets with similar symmetry properties include $FeBO_3$, $FeCO_3$, $CoCO_3$, $NiCO_3$, $MnBO_3$ [10,23].

In crystals with even lower symmetry, such as orthorhombic orthoferrites RFeO3 (R=Y, La, Pr, Nd etc) one can also find that $M_k = D_{kl}L_l$. Also here, a two-fold axis is the odd symmetry element. Although the antiferromagnetic Néel vector is not an axial vector, one of its components behaves similarly to a particular component of the magnetization vector under all symmetry operation allowed by the crystal structure.

It means that also these antiferromagnetic crystals allow to do a substitution $M_k \rightarrow M_k + D_{kl}L_l$ in Eq.7a and thus obtain an antiferromagnetic term in the magneto-optical Faraday effect.

For instance, in yttrium orthoferrite YFeO3, the spins of the $Fe^{3+}$ ions are ordered antiferromagnetically. Applying every single symmetry operation allowed by the crystal in the paramagnetic state, one finds that $M_x$ transforms similarly to $L_z$ and $M_z$ transforms similarly to $L_x$. Hence, we can write $M_z = D_{zx}L_x$ and $M_x = D_{xz}L_z$.

The situation is thus very comparable to that in hematite. Naturally, the similarity in magnetic structure motivated similar experiments. In particular, B. Krichevstov et al experimentally measured the ratio between the ferromagnetic and antiferromagnetic magneto-optical Faraday effect in YFeO3 [27]. The measurements were performed for light propagating along the $z$-axis and a magnetic field also along the z-axis. The term of the dielectric permittivity resulting in the magneto-optical Faraday effect can thus have 3 contributions

$$\varepsilon_{xy}^{(a)} = \chi_{xyz}^{(H)}H_z + \chi_{xyz}^{(M)}M_z + \chi_{xyx}^{(L)}L_x \tag{10}.$$

Performing the measurements as a function of the applied magnetic field, they found that the slope of the actually observed magneto-optical Faraday effect is 6 times smaller than the slope expected from the field-dependent and magnetization dependent contributions, taken together $\chi^H H_z + \chi^M M_z$ (see Fig. 3). Hence it was concluded that, similarly to the Hall effect in hematite α-Fe2O3, the magneto-optical Faraday effect in yttrium orthoferrite YFeO3 is dominated by the antiferromagnetic contribution $\chi_{xyx}^{(L)}L_x$.

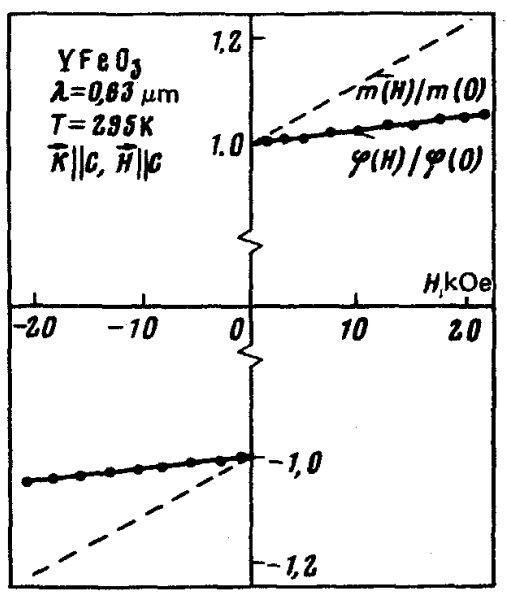

**Figure** 3.*Antiferromagnetic/altermagnetic magneto-optical Faraday effect in yttrium orthoferrite YFeO3 [27].*

In the same way, one can also expect additional contributions to $\varepsilon_{kl}^{(s)}$ (see Eq.7b) that are linear with respect to the antiferromagnetic Neel vector. In particular, substituting $M_k \to M_k + D_{kl}L_l$ in the $\chi_{klnp}^{(HM)} H_n M_p$-term, we obtain a new term in $\varepsilon_{kl}^{(s)}$

$$\varepsilon_{kl}^{(s)} \sim \chi_{klnp}^{(HM)} H_n L_p.$$

This term will result in magnetic linear birefringence, which will scale linearly with respect to one of the components of the antiferromagnetic Neel vector **L**. This effect was studied experimentally in Ref.[28].

Fluorides (CoF$_2$, MnF$_2$, FeF$_2$, NiF$_2$) as well as RuO$_2$ are also antiferromagnets with rotation as the odd symmetry element [29]. All these antiferromagnets have similar crystal structure in the paramagnetic state, described by the space group $D_{4h}^{14}$. The crystal structure and the ions corresponding to two antiferromagnetically coupled sublattices with magnetizations **M**$_1$ and **M**$_2$ are shown in Fig. 4.

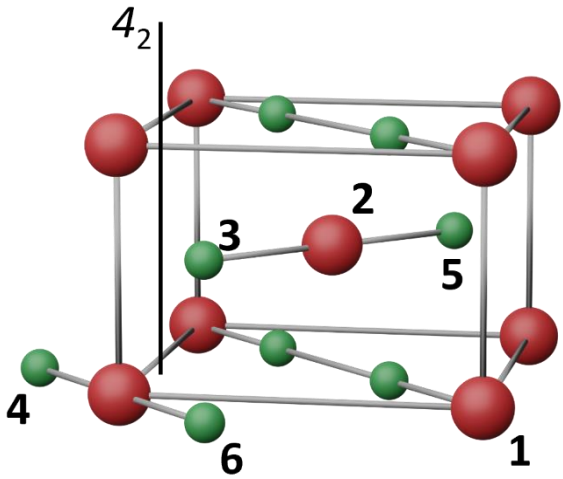

**Figure** 4. *Crystal structure of transition-metal fluorides MnF$_2$ and CoF$_2$. Red spheres are the positions of the magnetic ions. Numbers 1 and 2 indicate magnetic sublattices with magnetization **M**$_1$ and **M**$_2$, respectively. Green spheres (2,4,5,6) indicate the fluorine positions F$^-$. The odd symmetry element here is a screw axis 4$_2$,*

The magnetizations of the two sublattices are pointing along the *z*-axis. We define the antiferromagnetic Néel vector as

$$\mathbf{L} = \mathbf{M}_1 - \mathbf{M}_2$$

The odd symmetry element here is a screw axis 4$_2$, which means that a combination of rotation around the axis and a translation over ½ the unit cell parallel to the z-axis leaves the crystal unchanged (see Fig.4). It can be shown that the combination $L_x M_y + L_y M_x$ is invariant under every single symmetry operation allowed by the crystal structure [10,29]. Hence, we can also write that $M_y = D_{yx} L_x$ and $M_x = D_{xy} L_y$. In Eq.7a for $\varepsilon_{kl}^{(a)}$ we can thus substitute $M_y \to M_y + D_{yx} L_x$ and $M_x \to M_x + D_{xy} L_y$. In MnF$_2$ and CoF$_2$, for instance, the spins are aligned along the *z*-axis, in the ground state $L_x = L_y = 0$. Hence, although the Dzyaloshinskii-Morya interaction is allowed by symmetry, it does not result in a spin canting in the ground state.

However, applying external stimuli one can set another thermodynamic equilibrium for **L** or dynamically bring the Néel vector out of equilibrium such that $L_x \neq 0$ and/or $L_y \neq 0$. These components can contribute to $\varepsilon_{zx}^{(a)}$ and $\varepsilon_{zy}^{(a)}$, respectively. In particular, such a contribution to $\varepsilon_{zx}^{(a)}$ will result in an antiferromagnetic Faraday effect linear with respect to $L_x$ for light propagating along the *y*-axis.

**In antiferromagnets, where the odd symmetry element is a, proper or improper, rotation (as in altermagnets), read-out of antiferromagnetic domains with mutually opposite Néel vectors can become possible if the crystal symmetry allows to state that**

$$M_k \sim D_{kl} L_l,$$

**where $D_{kl}$ is a phenomenological tensor similar to the tensor of the Dzyaloshinskii-Moryia interaction.**

The crystal symmetry of fluorides ($CoF_2$, $MnF_2$, $NiF_2$, $FeF_2$) and $RuO_2$ also allows to induce mutual correlations between **M** and **L** similar to those due to the Dzyaloshinskii-Moryia interaction. A mechanical strain $\Sigma_{xy}$, for instance, induces in these compounds a magnetization along the *z*-axis. This piezomagnetic effect was studied in $CoF_2$ by A. S. Borovik-Romanov [30], being described phenomenologically as $M_z = \eta_{zzxy}^{(\Sigma)} L_z \Sigma_{xy}$. As the strain $\Sigma_{xy}$ has the very same symmetry as $H_x H_y$ and $E_x E_y$, one can predict that in these materials a magnetization along the *z*-axis can be induced by external stimuli: $M_z = \eta_{zzxy}^{(H)} L_z H_x H_y$ and $M_z = \eta_{zzxy}^{(E)} L_z E_x E_y$.

Consequently, in Eq.7a for $\varepsilon_{kl}^{(a)}$ we can substitute $M_z \to M_z + \eta_{zzxy}^{(\Sigma)} L_z \Sigma_{xy}$, $M_z \to M_z + \eta_{zzxy}^{(H)} L_z H_x H_y$, $M_z \to M_z + \eta_{zzxy}^{(E)} L_z E_x E_y$. Thus, for light propagating along the z-axis we can expect a magneto-optical Faraday effect linear with respect to the *z*-component of the antiferromagnetic Neel vector $L_z$. Here $\eta_{zzxy}^{(\Sigma)}$, $\eta_{zzxy}^{(H)}$, $\eta_{zzxy}^{(E)}$ are components of phenomenological tensors that describe the corresponding effect.

*4C. Antiferromagnets with space inversion as odd symmetry element*

The crystal structure of $Cr_2O_3$ is absolutely identical to that of hematite. However, a different alignment of spins of the magnetic $Cr^{3+}$ ions, compared to $Fe^{3+}$ in hematite, appears to result in dramatically different magnetic properties. The spins of $Cr^{3+}$ are aligned along the $3_z$ axis, but such that $\mathbf{M}_1$ and $\mathbf{M}_3$ are mutually parallel, while $\mathbf{M}_2$ and $\mathbf{M}_4$ are also mutually parallel but pointing in the opposite direction. The antiferromagnetic Néel vector in $Cr_2O_3$ is thus defined as

$$\mathbf{L} = \mathbf{M}_1 - \mathbf{M}_2 + \mathbf{M}_3 - \mathbf{M}_4$$

The odd symmetry element for this antiferromagnet is space inversion. It means that one can construct the following invariants with respect to the symmetry operations allowed by the crystal: $L_z M_y E_y$, $L_z M_x E_x$, $L_z M_z E_z$, where $E_x, E_y, E_z$ are the components of an applied electric field or any other vector having the symmetry of an electric field. It means that applying the electric field along the *y*-axis, we can expect $M_y = D_{yz}(E) L_z$. Similarly, applying the electric field along the *x*-axis gives $M_x = D_{xz}(E) L_z$ and if the electric field is along the *z*-axis, we have $M_z = D_{zz}(E) L_z$. Hence, the equations show that an electric field induces a magnetization in the material and for the antisymmetric part of the dielectric permittivity in $Cr_2O_3$ we can write

$$\varepsilon_{xy}^{(a)} = \chi_{xyzz}^{LE} E_z L_z \qquad (11a),$$

$$\varepsilon_{xz}^{(a)} = \chi_{xzyz}^{LE} E_y L_z \qquad (11b),$$

$$\varepsilon_{yz}^{(a)} = \chi_{yzxz}^{LE} E_x L_z \qquad (11c).$$

These equations correspond to circular birefringence and dichroism in $Cr_2O_3$ induced by an external electric field [31-35]. This effect scales linearly with the antiferromagnetic Néel vector and thus facilitates visualization of 180-degrees antiferromagnetic domains in this material, as shown in Fig. 5.

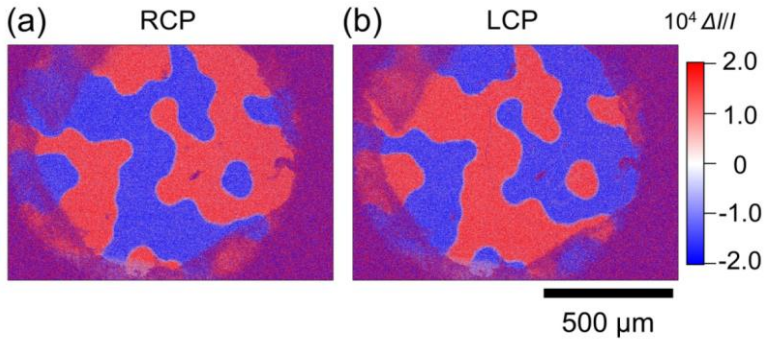

**Figure** 5. *Visualization of antiferromagnetic domains in an antiferromagnetic $Cr_2O_3$ (001) crystal. Two-dimensional maps of the electric field-induced magnetic circular dichroism obtained with right (a) and left (b) circularly polarized monochromatic light. The photon energy is equal to 2.34 eV. The images were taken at 83 K [35].*

The role of the electric field can be played by any other vector with the symmetry of an electric field. For instance, due to inversion symmetry breaking at surfaces of $Cr_2O_3$, the role of the electric field can be played by an electric polarization that emerges at every surface and is directed along the surface normal. Circular dichroism and birefringence at surfaces of $Cr_2O_3$ can only be observed in reflection [36] and are absent in transmission, because the effects at two surfaces are supposed to compensate one another. The effects also can be interpreted in terms of axions hosted by antiferromagnetic surfaces and described in terms optical axion electrodynamics [37].

Hence, we conclude

**In antiferromagnets, for which space inversion is the odd symmetry element, magneto-optical effects linear in L can be induced by external stimuli described by polar vectors, such as an electric field or current. The magneto-optical effects are also allowed upon reflection of light from crystal surfaces such that the role of the polar vector is played by the sample normal.**

### 5. Spin dynamics in antiferromagnets

We again consider the simplest case of an antiferromagnet with just two magnetic sublattices having magnetizations $\mathbf{M}_1$ and $\mathbf{M}_2$, respectively. For each of the sublattices we can write, in the non-dissipative approximation, the Landau-Lifshitz equation

$$\frac{\partial \mathbf{M}_1}{\partial t} = -\gamma\left[\mathbf{M}_1 \times \mathbf{H}_1^{(\text{eff})}\right] \qquad (12a)$$

and

$$\frac{\partial \mathbf{M}_2}{\partial t} = -\gamma\left[\mathbf{M}_2 \times \mathbf{H}_2^{(\text{eff})}\right] \qquad (12b),$$

where $\gamma$ is the gyromagnetic ratio that is assumed to be the same for the two sublattices. $\mathbf{H}_1^{(\text{eff})} = -\frac{1}{\mu_0}\frac{\partial \Phi}{\partial \mathbf{M}_1}$ and $\mathbf{H}_2^{(\text{eff})} = -\frac{1}{\mu_0}\frac{\partial \Phi}{\partial \mathbf{M}_2}$ are effective magnetic fields experienced by the first and the second sublattice, respectively, and $\Phi$ is the thermodynamic potential of the antiferromagnet written in terms of $\mathbf{M}_1$ and $\mathbf{M}_2$.

Since $\mathbf{M} \cdot \mathbf{L} = 0$ and $\mathbf{M}^2 + \mathbf{L}^2 = 4M_0^2$, where $M_0 = |\mathbf{M}_1| = |\mathbf{M}_2|$, we can rewrite the same equations with different terms:

$$\frac{\partial \mathbf{M}}{\partial t} = -\gamma\left[\mathbf{M} \times \mathbf{H}_M^{(\text{eff})}\right] - \gamma\left[\mathbf{L} \times \mathbf{H}_L^{(\text{eff})}\right] \qquad (13a)$$

and

$$\frac{\partial \mathbf{L}}{\partial t} = -\left[\mathbf{M} \times \mathbf{H}_L^{(\text{eff})}\right] - \left[\mathbf{L} \times \mathbf{H}_M^{(\text{eff})}\right] \qquad (13b).$$

where $\mathbf{H}_L^{(\text{eff})} = -\frac{1}{\mu_0}\frac{\partial \Phi}{\partial \mathbf{L}}$, $\mathbf{H}_M^{(\text{eff})} = -\frac{1}{\mu_0}\frac{\partial \Phi}{\partial \mathbf{M}}$ and $\Phi$ is the thermodynamic potential of the antiferromagnet written in terms of $\mathbf{L}$ and $\mathbf{M}$.

The art of constructing the thermodynamic potential is a cornerstone of this approach. When finding a correct form of the thermodynamic potential, one simply includes all possible terms which depend on $\mathbf{M}$ and $\mathbf{L}$ and which are invariant under all symmetry operations allowed by the crystal structure of the antiferromagnet. The form of the thermodynamic potential can be substantially simplified taking into account that $|\mathbf{M}| \ll |\mathbf{L}| \approx 2M_0$. Moreover, among the three invariants $\mathbf{L}^2$, $\mathbf{M}^2$ and $\mathbf{M} \cdot \mathbf{L}$, all representing the isotropic exchange interaction, one can leave in the potential only the one proportional to $\mathbf{M}^2$. Other terms proportional to $\mathbf{M}^2$ and representing weaker interactions can thus be neglected since $|\mathbf{M}| \ll |\mathbf{L}|$. Hence the thermodynamic potential per unit volume can be written as

$$F = \frac{\epsilon}{2}\mathbf{M}^2 + \frac{\alpha}{2}(\nabla \mathbf{L})^2 + W_a(\mathbf{L}) - D_{ik}M_i L_k - \mu_0 \mathbf{M} \cdot \mathbf{H} + W_{int}. \qquad (14)$$

The first two terms in this equation represent the exchange energy, where $\epsilon = \frac{2M_0}{H_{ex}}$ is the corresponding exchange constant, $H_{ex}$ is the effective magnetic field of the exchange interaction, $\alpha$ is a phenomenological parameter characterizing the strength of the inhomogeneous exchange interaction, $W_a(\mathbf{L})$ is the energy of magnetic anisotropy, the fourth term represents the Dzyaloshinskii-Moryia interaction, the fifth is the Zeeman energy and the last one represent light-spin interaction in the antiferromagnet due to opto-magnetic effects. Below, the parts of the effective fields corresponding to this $W_{int}$ will be called opto-magnetic field $\mathbf{H}_{M,L}^{(OM)}$ ($\mathbf{H}_L^{(OM)} = -\frac{1}{\mu_0}\frac{\partial W_{int}}{\partial \mathbf{L}}$ and $\mathbf{H}_M^{(OM)} = -\frac{1}{\mu_0}\frac{\partial W_{int}}{\partial \mathbf{M}}$).

Note that the way $\mathbf{H}_M^{(\text{eff})}$ acts on spins does depend on the magnetic structure of the antiferromagnet. Assuming that $|\mathbf{M}| \ll |\mathbf{L}| \approx 2M_0$, we further treat $\mathbf{M}$ as a "slave variable", which is expressed in terms of the antiferromagnetic Néel vector and its time derivative:

$$\mathbf{M} = \frac{2M_0}{H_{ex}}[\mathbf{H}_{\text{eff}} - \mathbf{l}(\mathbf{H}_{\text{eff}} \cdot \mathbf{l}) + \frac{1}{\gamma}(\frac{\partial \mathbf{l}}{\partial t} \times \mathbf{l})] \qquad (15),$$

where $\mathbf{l} = \mathbf{L}/|\mathbf{L}| \simeq \mathbf{L}/2M_0$, $\mathbf{H}_{\text{eff}} = \mathbf{H}_D(\mathbf{l}) + \mathbf{H}_M^{(\text{eff})}$, $\mathbf{H}_D$ is the effective field of the Dzyaloshinskii-Moryia interaction ($H_{D,i} = D_{ik}L_k/\mu_0$).

The Lagrangian in terms of the antiferromagnetic Néel vector $\mathbf{l}$ is thus

$$\mathcal{L} = \frac{M_0}{\gamma^2 H_{ex}}(\frac{\partial \mathbf{l}}{\partial t})^2 - W - \frac{2M_0}{\gamma H_{ex}}(\mathbf{H}_{\text{eff}} \cdot [\mathbf{l} \times \frac{\partial \mathbf{l}}{\partial t}]). \qquad (16),$$

where

$$W = \frac{A}{2}(\nabla \mathbf{l})^2 + W_a(\mathbf{l}) + \frac{M_0}{H_{ex}}[(\mathbf{H}_{\text{eff}} \cdot \mathbf{l})^2 - \mathbf{H}_{\text{eff}}^2] \qquad (17).$$

Derivations of the equations and further details of the sigma model can be found in Ref. 38.

The similarity of this Lagrangian with the Lagrangian for a particle moving with a velocity $d\mathbf{l}/dt$ and a coordinate $\mathbf{l}$ allows one to consider the problem of light-spin interaction as the problem of the action of an external force on a point particle. The effective fields caused by light-spin interaction, $\mathbf{H}_M^{(OM)}$ and $\mathbf{H}_L^{(OM)}$, contribute to the external force which drives spin dynamics; but the roles of the fields are different.

If $H_D(\mathbf{l}) = 0$, the role of the driving force in this case can be played by the time-derivative of the opto-magnetic field $\mathbf{H}_M^{(OM)}$. The antiferromagnetic Néel vector will move only during the action of the pulse. Hence, if the pulse is much shorter than the period of the spin resonance, such an excitation will give only a small deviation of the Néel vector from its equilibrium orientation. If $\mathbf{H}_D(\mathbf{l}) \neq 0$, the role of the driving force can be played directly by $\mathbf{H}_M^{(OM)}$. In this case, if the opto-magnetic field $\mathbf{H}_M^{(OM)}$ is a pulse with duration much shorter than the period of the spin resonance, the effective field acts impulsively. During the action of the pulse, the orientation of $\mathbf{l}$ hardly changes, but the spins acquire a momentum proportional to $d\mathbf{l}/dt$ and move long after the action of the pulse, as if the spins have inertia [39] (see Fig. 6). From Eq.17 it is also seen that terms proportional to $\mathbf{H}_D \cdot \mathbf{H}_M^{(OM)}$ or $(\mathbf{H}_D \cdot \mathbf{l})(\mathbf{H}_M^{(OM)} \cdot \mathbf{l})$ do trigger inertial spin dynamics.

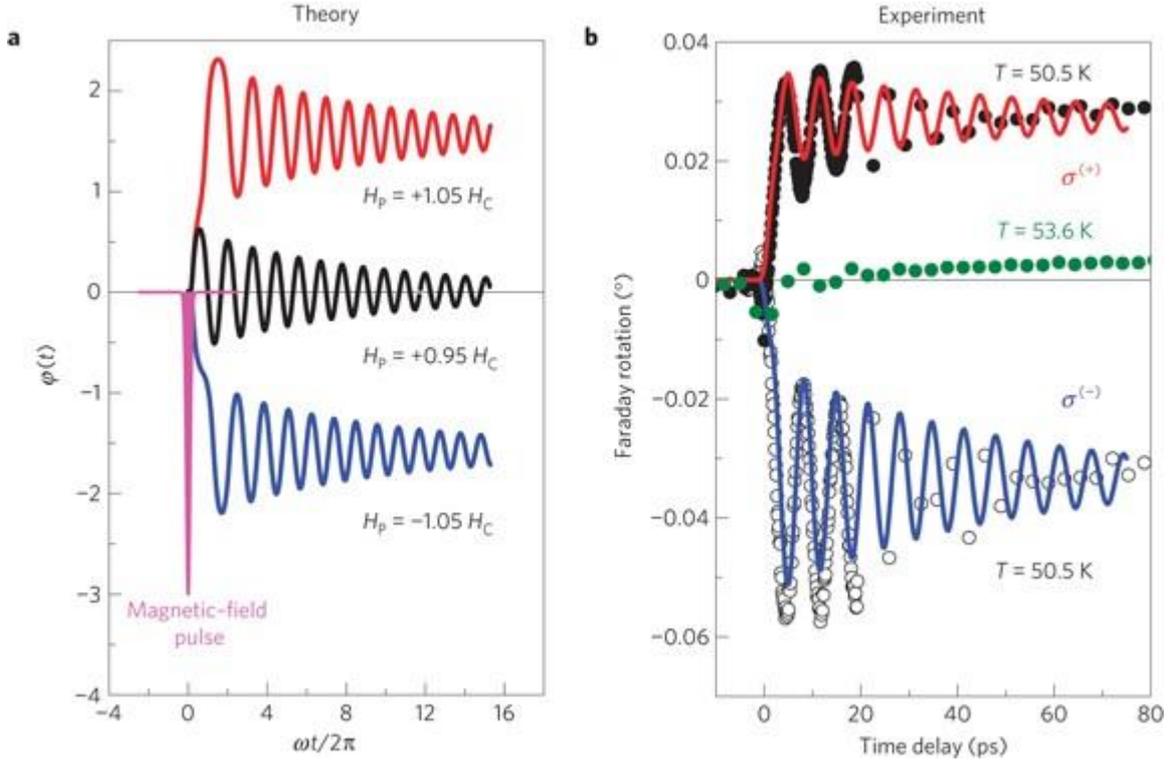

**Figure** 6. *Theoretical (panel a) and experimental (panel b) demonstration of inertia-driven spin switching in HoFeO3. In theory, the antiferromagnetic Néel vector **l** hardly moves during the action of the magnetic field pulse. Instead, the spins acquire a momentum proportional to $d\mathbf{l}/dt$ and reorient long after the pulse. The behavior is reproduced experimentally, where the effective magnetic field is generated by a circularly polarized femtosecond laser pulse [39].*

Referring to the classification of antiferromagnets introduced above, we will analyse $\mathbf{H}_L^{(OM)}$ and $\mathbf{H}_M^{(OM)}$ generated by light for each of the three classes.

Interestingly, the structure of the Landau-Lifshitz equations for antiferromagnets implies that even if $\mathbf{M} = 0$, $\mathbf{H}$ and $\mathbf{H}_M^{(OM)}$ can still excite spins by acting on the antiferromagnetic Néel vector (see Eq. 13b). The role of the driving force in this case is played by the time-derivatives of the magnetic fields [40,41]. Hence, due to the magnetic-dipole (Zeeman) term, a THz magnetic field can excite spins practically in any antiferromagnet, including those with translation symmetry as the odd element [41,42].

From Eq. 2 we can state that the part of the specific internal energy that depends on light is given by $W_{int} = \frac{1}{2}\varepsilon_0 \varepsilon_{kl}(\mathbf{H},\mathbf{M},\mathbf{L}) E_k(\omega) E_l^*(\omega)$. Hence, for the effective fields generated by light we have

$$\mathbf{H}_M^{(OM)} = -\frac{1}{\mu_0} \frac{\partial \varepsilon_{kl}(\mathbf{H},\mathbf{M},\mathbf{L})}{\partial \mathbf{M}} E_k(\omega) E_l^*(\omega) \tag{18}$$

and

$$\mathbf{H}_L^{(OM)} = -\frac{1}{\mu_0} \frac{\partial \varepsilon_{kl}(\mathbf{H},\mathbf{M},\mathbf{L})}{\partial \mathbf{L}} E_k(\omega) E_l^*(\omega) \tag{19}.$$

## 6. Opto-magnetism of antiferromagnets

*6A. Antiferromagnets with translation as odd symmetry element*

Even if the net magnetization is null, practically in any antiferromagnet we can expect that $\frac{\partial \varepsilon_{kl}}{\partial \mathbf{M}} \neq 0$ (see Eq. 7a). Hence, the effective field $\mathbf{H}_M^{(OM)}$ generated by a femtosecond laser pulse can excite spins due to the inverse, opto-magnetic, Faraday effect (see Fig. 7 and Ref. 41)

$$H_M^{(OM)}{}_n = -\frac{1}{\mu_0}\chi_{nkl}^{(M)} E_k(\omega) E_l^*(\omega) \quad (20).$$

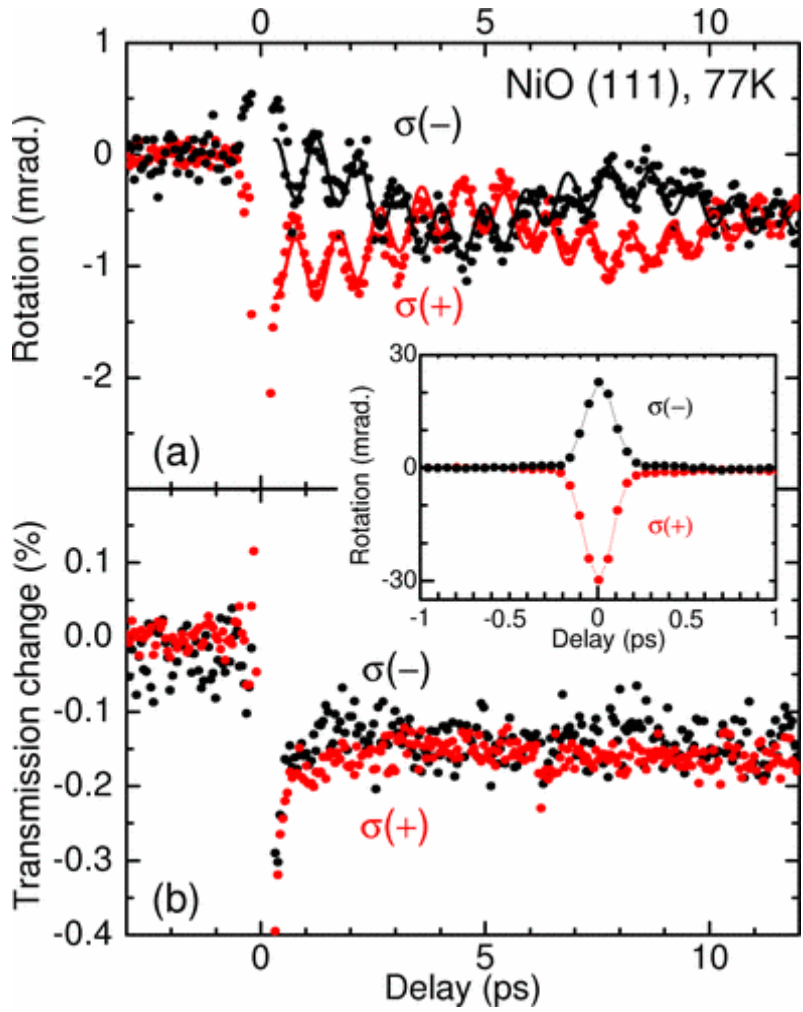

**Figure** 7. *Ultrafast laser-induced dynamics triggered due to ultrafast inverse Faraday effect in NiO (111). The dynamics is observed in time-resolved (a) polarization rotation and (b) transmission change in NiO (111) for pump helicities σ(+,−) [41].*

Similarly, from Eq. 7b we can expect excitation of spins via the inverse Cotton-Mouton effect [43]

$$H_L^{(OM)}{}_n = \chi^{(L)}_{nklp} E_k(\omega) E_l^*(\omega) L_p \tag{21}$$

*6B. Antiferromagnets with rotation axis as odd symmetry element*

Ultrafast opto-magnetic fields due to the inverse Faraday effect and the inverse Cotton-Mouton effect were first reported for this particular class of antiferromagnets [44-45]. Symmetry properties of this class appear to coincide with those in altermagnets, as explained in Refs.[7,17]. Hence, these antiferromagnets do allow other types of opto-magnetic effects. For instance, in $FeBO_3$ symmetry analysis predicts the existence of an opto-magnetic field $H_L^{(OM)}{}_x$ induced by circularly polarized light propagating in the direction of the *y*-axis of the crystal.

$$H_L^{(OM)}{}_x = \chi^{(L)}_{xxz} E_x(\omega) E_z^*(\omega) \tag{22}$$

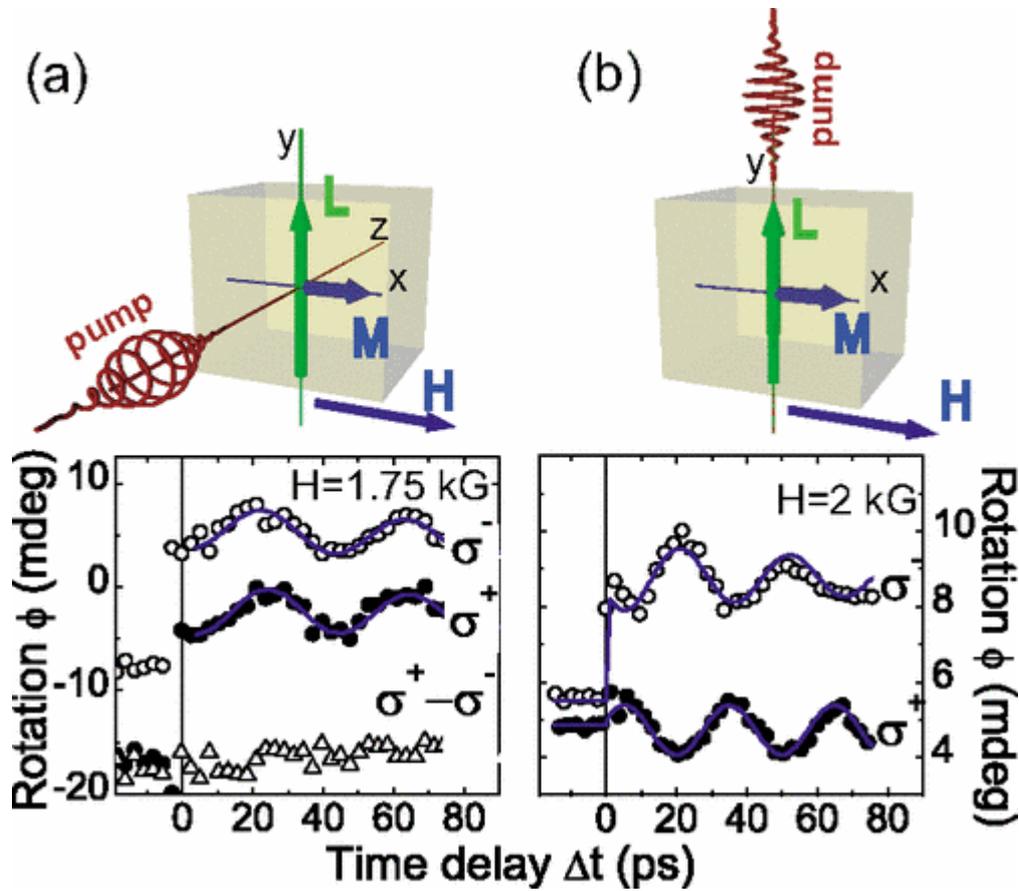

**Figure 8.** *Spin precession excited in $FeBO_3$ by circularly polarized pump pulses propagating along (a) the z-axis ($3_z$-axis in Fig. 1) and along (b) the y axis. $\sigma^+ - \sigma^-$ is the difference between the spin precession amplitude excited by right- and left-handed circularly polarized pump pulses [46].*

Experiments in this geometry were reported in Ref. [46]. In this experiment (see Fig. 8), however, light must propagate perpendicular to the main axis of the crystal, experiencing a substantial linear birefringence. It is questionable if it is correct to talk in this case about propagation of circularly polarized light, because a circularly polarized wave is no longer an eigen state here.

*6C. Antiferromagnets with space inversion as odd symmetry element*

Antiferromagnets of this symmetry are magneto-electric materials. For instance, in $Cr_2O_3$ symmetry analysis predicts that THz electric fields at a frequency $\Omega$ $\mathbf{E}(\Omega) = (E_x, E_y, E_z)$ can generate THz magnetic fields $\mathbf{H}(\Omega) = (\chi_{xxz}^{(ME)} E_x L_z, \chi_{yyz}^{(ME)} E_y L_z, \chi_{zzz}^{(ME)} E_z L_z)$, where $\chi_{kln}^{(ME)}$ are the corresponding elements of the magneto-electric tensor. Similar fields can also be induced by THz currents, as claimed in Ref. [47-49] for the case of $Mn_2Au$.

Extending further the concept of opto-magnetic fields for this type of antiferromagnets, symmetry analysis also predicts effective fields induced by circularly polarized light

$$H_L^{(OM)}{}_z = \chi_{zxyz}^{(LE)} E_x(\omega) E_y^*(\omega) E_z(0) + \chi_{zxzy}^{(LE)} E_x(\omega) E_z^*(\omega) E_y(0) + \chi_{zyzx}^{(LE)} E_y(\omega) E_z^*(\omega) E_x(0).$$

Such an opto-magnetic field has been recently reported in Ref. 37. In particular, it was reported that circularly polarized light can control spins in the fully compensated two-dimensional even-layered antiferromagnetic $MnBi_2Te_4$. It was shown that the mechanism of the control has the very same origin as an antiferromagnetic circular dichroism, which appears only in reflection but is absent in transmission. It was argued that the optical control and circular dichroism both arise from the optical axion electrodynamics.

# 7. Conclusions

Observation of a magneto-optical Faraday effect in a material having antiferromagnetic spin order and vanishingly small or even zero net magnetization, is presently seen as a signature of altermagnetism [7]. In this review we formulate the main principles allowing to predict if light can visualize 180-degrees antiferromagnetic domains via the magneto-optical Faraday effect or any other magneto-optical effect, which is linearly proportional to the antiferromagnetic Néel vector. We show that magneto-optical effects in altermagnets, but also in other types of materials with antiferromagnetic ordering of spins, can be predicted, employing the principles of thermodynamics and symmetry analysis of antiferromagnets pioneered by Dzyaloshinksii, without any calculations of the electronic structure and even without introduction of the term of "altermagnetism". Hence, magneto-optics does not seem to be the best tool to reveal the counter-intuitive physics originating from the intriguing peculiarities of the electronic structure of altermagnets. Following the same principles, we extended the symmetry analysis on optically generated effective magnetic fields that can excite antiferromagnetically coupled spins in materials known as altermagnets and antiferromagnets, in general. The predictions confirmed by a plethora of experimental observations reveal multiple mechanisms to excite spins with the help of light, practically for every type of antiferromagnets.

## Acknowledgements

AVK is grateful to J. Sinova, T. Jungwirth, L. Šmejkal, S. Cheong as well as to all participants of the SPICE workshop "Altermagnetism: emerging opportunities in a new magnetic phase" for stimulating discussions, which triggered the idea to write this review. The authors would like to thank T. Gareev for his help with figures in the paper. The work was supported by the European Research Council ERC (Grant Agreement No. 101054664 (SPARTACUS) and Grant Agreement No.856538 (3D-MAGiC)).

# REFERENCES

[1] C. Kittel, *Introduction to solid state physics* (8th ed., John Wiley & Sons, 2004).

[2] J. Nogués, I. K. Schuller, Exchange bias, *Journal of Magnetism and Magnetic Materials,* **192**, 203–232 (1999).

[3] T. Jungwirth, X. Marti, P. Wadley, and J. Wunderlich, Antiferromagnetic spintronics. *Nature Nano.* **11**, 231–241 (2016).

[4] V. Baltz, A. Manchon, M. Tsoi, T. Moriyama, T. Ono, and Y. Tserkovnyak, Antiferromagnetic spintronics, *Rev. Mod. Phys.* **90**, 015005 (2018).

[5] A. Barman et al, The 2021 magnonics roadmap, *J. Phys.: Condens. Matter* **33**, 413001 (2021).

[6] P. Němec, M. Fiebig, T. Kampfrath, and A. V. Kimel, Antiferromagnetic opto-spintronics. *Nat. Phys.* **14**, 229–241 (2018).

[7] L. Šmejkal, J. Sinova, and T. Jungwirth, Emerging research landscape of altermagnetism, *Phys. Rev. X* **12**, 040501 (2022).

[8] E. A. Turov, Physical *Properties of magnetically ordered crystals*, Academic Press, New York (1965).

[9] E. A. Turov, A. V. Kolchanov, V. V. Men'shenin, I. F. Mirsaev, V. V. Nikolaev, Magnetodynamics of antiferromagnets, *Phys. Usp*. **41**, 1191–1198 (1998).

[10] E. A. Turov, A. V. Kolchanov, V. V. Men'shenin, I. F. Mirsaev, and V. V. Nikolaev, *Symmetry and the physical properties of antiferromagnets* [in Russian], Fizmatlit, Moscow (2001). E. A. Turov, *Symmetry and physical properties of antiferromagnetics*, Cambridge International Science Publishing (2010).

[11] J. Stöhr and H. C. Siegman, *Magnetism: from fundamentals to nanoscale dynamics* (Springer Series in Solid-State Sciences) (Berlin: Springer) (2006).

[12] L. D. Landau, and E. M. Lifshitz, *Electrodynamics of continuous media*, Pergamon, New York, (1984).

[13] A. K. Zvezdin and V. A. Kotov, *Modern magnetooptics and magnetooptical materials*, CRC Press, Boca Raton (1997).

[14] S.-W. Cheong & F.-T. Huang, Trompe L'oeil Ferromagnetism. *npj Quant. Mater.* **5**, 37 (2023).

[15] S.-W. Cheong & F.-T. Huang, Altermagnetism with non-collinear spins, *npj Quantum Mater.* **9**, 13 (2024).

[16] M. Fiebig, *Nonlinear optics on ferroic materials*, John Wiley & Sons (2023).

[17] J. Krempaský *et al.* Altermagnetic lifting of Kramers spin degeneracy. *Nature* **626**, 517–522 (2024).

[18] V. G. Baryakhtar, Phase diagram of chromium in an external magnetic field, *JETP Letters* **30**, 619-622 (1979).

[19] L. D. Landau, and E. M. Lifshitz, *Theory of elasticity*, Pergamon, New York, (1984).

[20] I. E. Dzyaloshinksii, Domains and disclinations in antiferromagnets, *JETP Letters* **25**, 8 (1977).

[21] B. A. Ivanov, Mesoscopic antiferromagnets: statics, dynamics, and quantum tunneling (Review), *Low Temp. Phys.* **31**, 635–667 (2005).

[22] M. Kleiber, M. Bode, R. Ravlic, and R. Wiesendanger, Topology-induced spin frustrations at the Cr(001) surface studied by spin-polarized scanning tunneling spectroscopy, *Phys. Rev. Lett.* **85**, 4606 (2000).

[23] I. E. Dzialoshinskii, Thermodynamic theory of "weak" ferromagnetism in antiferromagnetic substances, *Sov. Phys. JETP* **5**, 1259-1272 (1957).

[24] I. Dzyaloshinsky, A thermodynamic theory of "weak" ferromagnetism of antiferromagnetics, *Journal of Physics and Chemistry of Solids* **4**, 241-255m (1958).
[25] K. B. Vlasov, V. N. Novogrudskiy, E. A. Rozenberg, *Hall effect in weak-ferromagnetic natural single crystal of hematite* [in Russian], Fizika Metallov I Metallovedenie **42**, 513-517 (1976).
[26] K. B. Vlasov et al. *Hall effect in hematite-single crystal at Morin transition*, Solid State Physics **22**, 1656-1659 (1980).
[27] B. B. Krichevtsov, K. M. Mukimov, R. V. Pisarev, M. M. Ruvinshtejn, Antiferromagnetic and ferromagnetic Faraday effect in the YFeO3 yttrium orthoferrite. JETP Letters **34**, 379-382 (1981).
[28 V.S. Merkulov, E.G. Rudashevskii, H. Le Gall, C. Leycuras, Linear magnetic birefringence of hematite in the vicinity of the Morin temperature, *Sov . Phys. JETP* **53**, 81 (1981).
[29] I. E. Kzialoshinskii, The magnetic structure of fluorides of the transition metals, Sov. Phys. JETP 6, 1120-1122 (1958). I. E. Dzialoshinskii, Zhurnal Eksperimentalnoy I Teoreticheskoy Fiziki **33**, 1454-1456 (1957) [in Russian].
[30] A. S. Borovik-Romanov, Piezomagnetism in the antiferromagnetic fluorides of Cobalt and Manganese, *Sov. Phys. JETP* **11**, 786- 793 (1960).
[31] T. H. O'Dell and E. A. D. White, Electric field induced Faraday rotation in chromic oxide, *Philos. Mag. A: J. Theor. Exp. Appl. Phys*. **22**, 649 (1970).
[32] B. B. Krichevtsov, V. V. Pavlov, and R. V. Pisarev, Nonreciprocal rotation of the polarization plane of light in the antiferromagnet Cr2O3 which is linear and quadratic in the electric field, Pis'ma Zh. Eksp. Teor. Fiz. **44**, 471 (1986) [*Sov. J. Exp. Theor. Phys. Lett*. **44**, 607 (1986)].
[33] B. B. Krichevtsov, V. V. Pavlov, and R. V. Pisarev, Nonreciprocal optical effects in antiferromagnetic Cr2O3 subjected to electric and magnetic fields, Zh. Eksp. Teor. Fiz. **94**, 284 (1988) [*Sov. Phys. JETP* **67**, 378 (1988)].
[34] J. Wang and C. Binek, Dispersion of Electric-Field-Induced Faraday Effect in Magnetoelectric Cr2O3, *Phys. Rev. Appl*. **5**, 031001 (2016).
[35] T. Hayashida, K. Arakawa, T. Oshima, K. Kimura, and T. Kimura, Observation of antiferromagnetic domains in Cr2O3 using nonreciprocal optical effects, *Phys. Rev. Research* **4**, 043063 (2022).
[36] B. B. Krichevtsov, V. V. Pavlov, R. V. Pisarev and V. N. Gridnev, Spontaneous non-reciprocal reflection of light from antiferromagnetic Cr2O3, *J. Phys.: Condens. Matter* **5** 8233 (1993).
[37] J.-X. Qiu et al, Axion optical induction of antiferromagnetic order, Nature Materials **22**, 583–590 (2023).
[38] B. A. Ivanov, Spin Dynamics for Antiferromagnets and Ultrafast Spintronics, *J. Exp. Theor. Phys.* **131**, 95–112 (2020).
[39] A. V. Kimel, B. A. Ivanov, R. V. Pisarev, P. A. Usachev, A. Kirilyuk & Th. Rasing, Inertia-driven spin switching in antiferromagnets, *Nature Physics* **5**, 727–731 (2009).
[40] K. Grishunin, E. A. Mashkovich, A. V. Kimel, A. M. Balbashov, and A. K. Zvezdin, Excitation and detection of terahertz coherent spin waves in antiferromagnetic α−Fe2O3, *Phys. Rev. B* **104**, 024419 (2021).
[41] T. Satoh, S.-J. Cho, R. Iida, T. Shimura, K. Kuroda, H. Ueda, Y. Ueda, B. A. Ivanov, F. Nori, and M. Fiebig, Spin oscillations in antiferromagnetic NiO triggered by circularly polarized light, *Phys. Rev. Lett*. **105**, 077402 (2010).
[42] T. Kampfrath et al, Coherent terahertz control of antiferromagnetic spin waves. *Nature Photonics* **5**, 31-34 (2011).

[43] C. Tzschaschel et al, Ultrafast optical excitation of coherent magnons in antiferromagnetic NiO, *Phys. Rev. B* **95**, 174407 (2017).

[44] A. V. Kimel, A. Kirilyuk, P. A. Usachev, R. V. Pisarev, A. M. Balbashov, R. V. Pisarev, and Th. Rasing, Ultrafast non-thermal control of magnetization by instantaneous photomagnetic pulses, *Nature* **435** 655 (2005).

[45] A. M. Kalashnikova, A. V. Kimel, R. V. Pisarev, V. N. Gridnev, A. Kirilyuk & Th. Rasing, Impulsive generation of coherent magnons by linearly polarized light in the easy-plane antiferromagnet $FeBO_3$, *Phys. Rev. Lett.* **99** 167205 (2007).

[46] A. M. Kalashnikova, A. V. Kimel, R. V. Pisarev, V. N. Gridnev, P. A. Usachev, A. Kirilyuk, Th. Rasing, Impulsive excitation of coherent magnons and phonons by subpicosecond laser pulses in the weak ferromagnet FeBO3, *Phys. Rev. B* **78** 104301 (2008).

[47] Y. Behovitset al. Terahertz Néel spin-orbit torques drive nonlinear magnon dynamics in antiferromagnetic Mn2Au, *Nature Communications* **14**, 6038 (2023).

[48] F. Freimuth, S. Bluegel, and Y. Mokrousov, Laser-induced torques in metallic antiferromagnets, *Phys. Rev. B* **103**, 174429 (2021).

[49] J. L. Ross, P-I. Gavriloaea, F. Freimuth, T. Adamantopoulos, Y. Mokrousov, R.F.L. Evans, R. Chantrell, R.M. Otxoa, O. Chubykalo-Fesenko, Antiferromagnetic switching in Mn2Au using a novel laser induced optical torque on ultrafast timescales*, arXiv:2311.00155* (2023).